\documentclass[5p]{elsarticle}

\usepackage{amssymb}

\journal{TIPP09 Proceedings in NIMA}

\begin{document}

\begin{frontmatter}



\title{Development of transparent silica aerogel over a wide range of densities}

%
%
%
%
%
\author[First,Second]{Makoto Tabata\corref{cor1}}
\ead{makoto@hepburn.s.chiba-u.ac.jp}
\cortext[cor1]{Corresponding author.} 
\author[Third]{Ichiro Adachi}
\author[First]{Yoshikazu Ishii}
\author[First]{Hideyuki Kawai}
\author[Fourth]{Takayuki Sumiyoshi}
\author[Fifth]{Hiroshi Yokogawa}
%
%
%
\address[First]{Department of Physics, Chiba University, Chiba, Japan}
\address[Second]{Institute of Space and Astronautical Science (ISAS), Japan Aerospace Exploration Agency (JAXA), Sagamihara, Japan}
\address[Third]{Institute of Particle and Nuclear Studies (IPNS), High Energy Accelerator Research Organization (KEK), Tsukuba, Japan}
\address[Fourth]{Department of Physics, Tokyo Metropolitan University, Hachioji, Japan}
\address[Fifth]{Advanced Materials Development Department, Panasonic Electric Works Co.,Ltd., Kadoma, Japan}

%
%
%
%
%
\begin{abstract}

We have succeeded in developing hydrophobic silica aerogels over a wide range of densities (i.e. refractive indices). A pinhole drying method was invented to make possible producing highly transparent aerogels with entirely new region of refractive indices of 1.06-1.26. Obtained aerogels are more transparent than conventional ones, and the refractive index is well controlled in the pinhole drying process. A test beam experiment was carried out in order to evaluate the performance of the pinhole-dried aerogels as a Cherenkov radiator. A clear Cherenkov ring was successfully observed by a ring imaging Cherenkov counter. We also developed monolithic and hydrophobic aerogels with a density of 0.01 g/cm$^3$ (a low refractive index of 1.0026) as a cosmic dust capturer for the first time. Consequently, aerogels with any refractive indices between 1.0026 and 1.26 can be produced freely.

\end{abstract}

%
%
%
%
%
%

\end{frontmatter}


%
%
%
%
%
%

\section{Introduction}

Silica aerogel is widely used as a Cherenkov radiator in high energy and nuclear experiments. We developed aerogel production technique for the Belle aerogel Cherenkov counter in 1990s, and have studied for its upgrade program since 2002 \cite{example1}. In our conventional production method \cite{example2}, we are able to produce hydrophobic aerogels for a refractive index range of 1.005-1.07 by using alcohol solvent for the alcogel (wet gel) synthesis. Introduction of new solvent, N,N-dimethylformamide (DMF), extended the upper limit of refractive index up to 1.11. Furthermore, the range of refractive index was extended again by a new production method in a recent study \cite{example3}. The production method also generates more transparent aerogels than conventional ones for a refractive index range of 1.06-1.26. The aerogels with high refractive index and high transparency will expand future possibilities of particle identification.

In addition, aerogel with a refractive index under 1.005 was also developed in recent years. It is a material with an extremely low density under 0.02 g/cm$^3$ because refractive index ($n$) of aerogel is expressed as a function of its density ($\rho $): $n = 1+\alpha \rho $, where $\alpha $ is a constant. It is expected that the low density aerogels will be utilized in cosmic dust sample return experiments, for example, the TANPOPO \cite{example4}, which is an astrobiology mission planned on the Japanese Experiment Module (JEM) in the International Space Station (ISS). 

In this article, the development of extremely low density aerogels is described in the next section. Section 3 is devoted to the introduction of the new production method for high refractive index aerogels. In Section 4, the transparency of new aerogels is explained. Then, conclusion is given after a description of a result of a beam test in Section 5.

\section{Extremely low density aerogel}

\begin{figure}[hbt]
\begin{center}
\includegraphics*[scale=0.40]{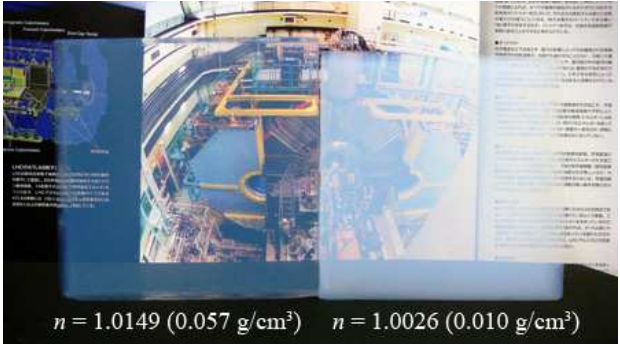}
\end{center}
\caption{\label{fig:fig1}
Photograph of low density aerogel tiles. Each of the dimensions is about $11\times 11\times 2$ cm$^3$.
}
\end{figure}

The lower limit of refractive index was 1.005 in our conventional production method so far. In a recent study for the development of a cosmic dust capture medium, we have succeeded in developing aerogel with a refractive index of 1.0026 (0.01 g/cm$^3$ in density). That was achieved by optimizing the mixing ratio of raw chemical solutions. A plastic mold for the alcogel synthesis allows us to generate well formed aerogel tiles. Fig. \ref{fig:fig1} shows a conventional aerogel with $n$ = 1.0149 (left) and the new aerogel with $n$ = 1.0026 (right). The transmission lengths at 400 nm wavelength are measured to be 26 mm and 6 mm, respectively. In the synthesis of the lowest density aerogels, the gelation process proceeds very slowly. As a result, the nanoscale structure of aerogels would be built up to have low transmittance at optical wavelengths.

%
%
%
%

Low density aerogel, a transparent solid, is the most suitable medium as a cosmic dust capturer. Aerogels enable intact capture of cosmic dust particles even at hypervelocity impact of a few kilometers per second. We think that lower density aerogels cause less thermal damage to incident particles. It is very important in order to carry out mineralogical and biochemical analyses after returning to the ground. The aerogels with a density of 0.01 g/cm$^3$ is transparent enough to find captured micron-size particles. A hypervelocity impact experiment was performed by using a two-stage light-gas gun at the Institute of Space and Astronautical Science, Japan Aerospace Exploration Agency (ISAS/JAXA) in 2007. Simulated cosmic dust particles with a diameter of 50 $\mu $m were successfully captured by our new 0.01 g/cm$^3$ aerogel blocks at a velocity of 4 km/s.

On the other hand, this low refractive index aerogels were tested as a Cherenkov radiator by using positron beams at the Laboratory of Nuclear Science (LNS) of Tohoku University in 2008 \cite{example5}. Although the number of detected photons ($N_{pe}$) was low, it was confirmed that Cherenkov light was emitted from the aerogels with $n$ = 1.0026. If we devise the optical system of a threshold Cherenkov counter, the low refractive index aerogels can be also used as a radiator.

%
%
%
%
\begin{figure}[t]
\begin{center}
\includegraphics*[scale=0.30]{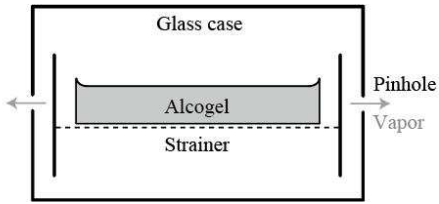}
\end{center}
\caption{\label{fig:fig2}
Schematic view of the standard pinhole-container.
}
\end{figure}
%

%
%
%
%
\begin{figure}[t]
\begin{center}
\includegraphics*[scale=0.20]{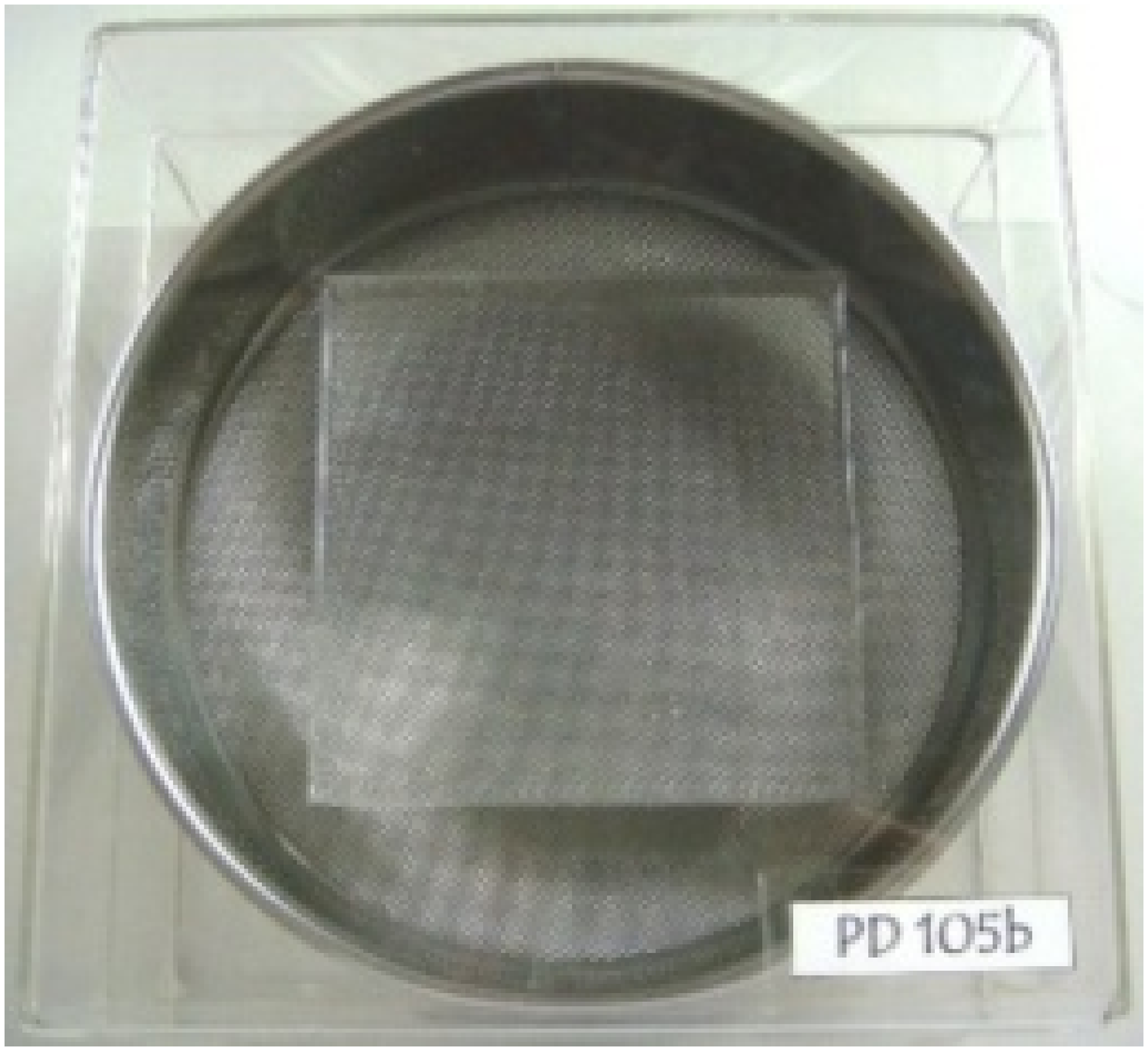}
\includegraphics*[scale=0.20]{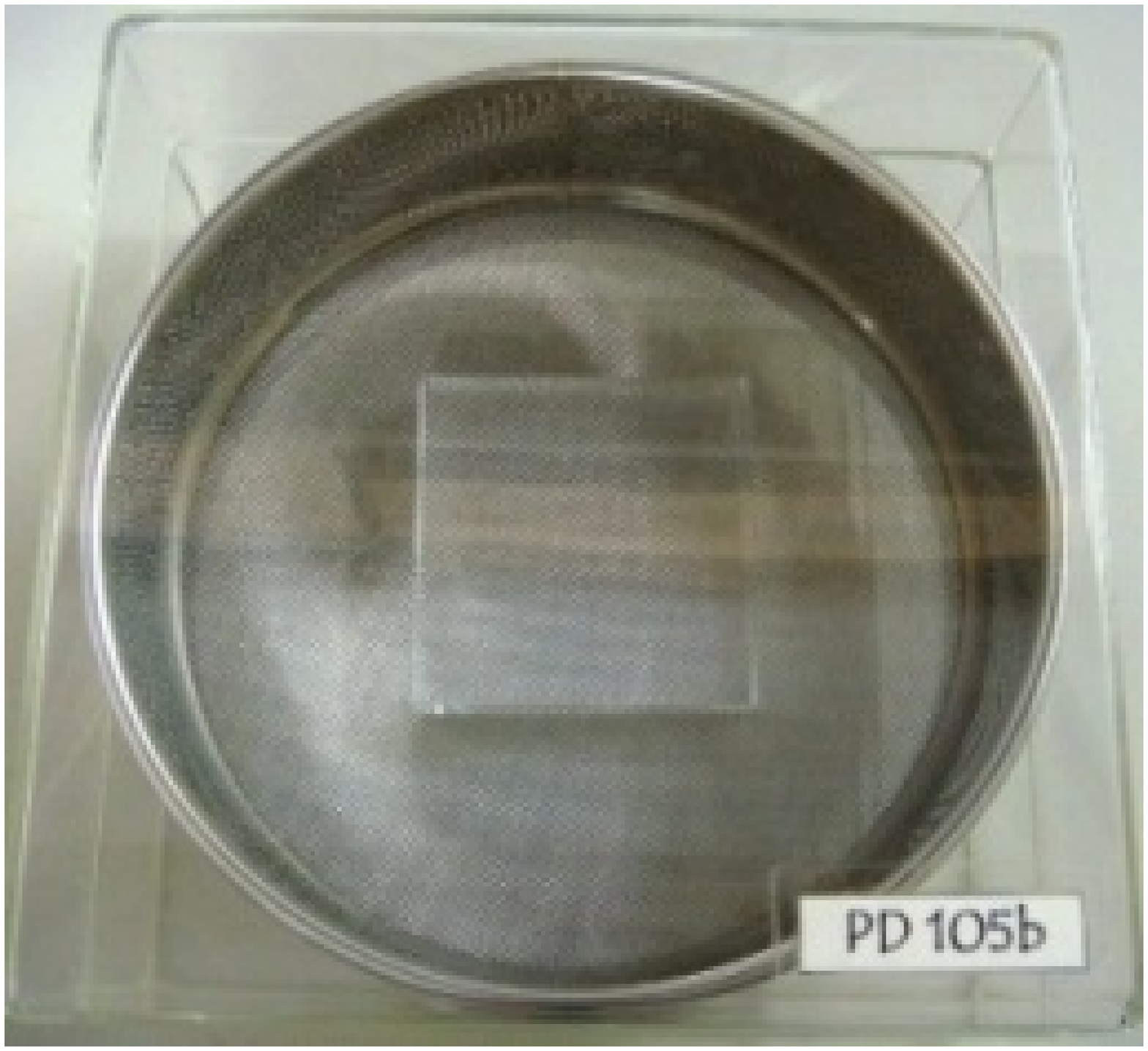}
\end{center}
\caption{\label{fig:fig3}
Photograph of an alcogel before and after the pinhole drying.
}
\end{figure}
%

%
%
%
%
\begin{figure}[t]
\begin{center}
\includegraphics*[scale=0.35]{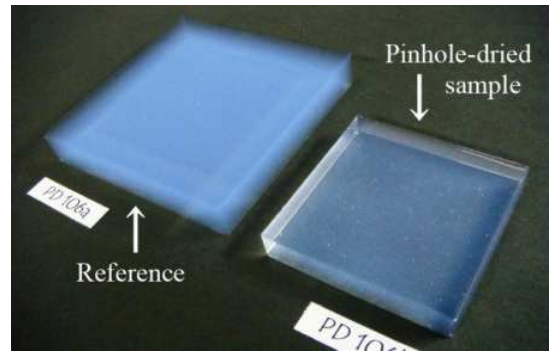}
\end{center}
\caption{\label{fig:fig4}
Photograph of a pair of pinhole-dried aerogel and reference (original non-shrink aerogel). The dimensions of the pinhole-dried aerogel are $6.5\times 6.5\times 1$ cm$^3$.
}
\end{figure}

\section{Pinhole drying method}

Pinhole drying method is a new and powerful technique to produce aerogels with high transparency and/or high refractive index. In our conventional method, refractive index of aerogels is determined in the alcogel synthesis process. As for the pinhole drying method, synthesized alcogels are dried to shrink in a pinhole-container as shown in Fig. \ref{fig:fig2}. The shrinkage of the alcogel generates highly transparent aerogel with higher refractive indices (densities) than the original aerogel. The shrinkage rate is well controlled by adjusting the design of pinhole-containers and room temperature to avoid any cracks. The pinhole drying period is determined by monitoring the weight of the alcogel to obtain a desired refractive index. In this method, we are able to obtain aerogels with $n$ = 1.06-1.26 after the supercritical drying. The dispersions of the refractive index is controlled among different tiles within $\pm $0.002 and $\pm $0.005 at $n$ = 1.10 and 1.20, respectively.

The pinhole drying process in the standard pinhole-container is shown in Fig. \ref{fig:fig3}. During several weeks or months, alcogels shrink keeping the original shape. Fig. \ref{fig:fig4} shows a typical pinhole-dried aerogel sample with a reference (original non-shrink sample). The refractive index increased from 1.067 to 1.198, and the transmission length at 400 nm wavelength also increased from 32 mm to 39 mm by the pinhole drying.

\section{Transparency}

\begin{figure}[t]
\begin{center}
\includegraphics*[scale=0.35]{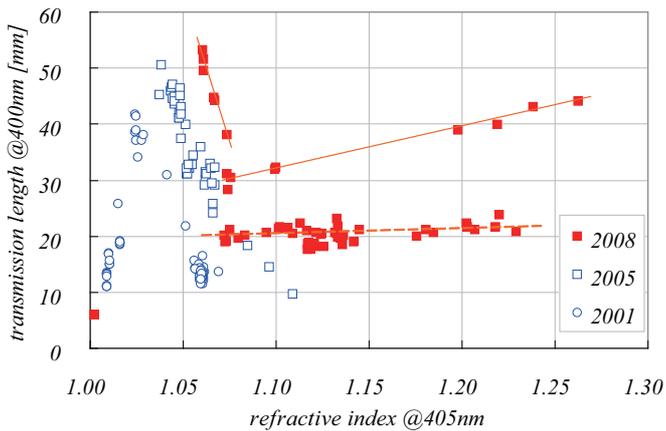}
\end{center}
\caption{\label{fig:fig5}
Transmission length as a function of refractive index.
}
\end{figure}

Fig. \ref{fig:fig5} shows transmission length measured at 400 nm wavelength as a function of refractive index of aerogels which can be produced by our current production methods. The open circles indicate aerogels by the conventional method in 1990s. As shown by the open squares, the significant improvement of transparency in the middle range of refractive index was accomplished by introducing DMF solvent to the alcogel synthesis process in 2005 \cite{example6}. At present, we are able to produce aerogels with $n$ = 1.0026-1.26. The closed squares indicate pinhole-dried aerogels except for the lowest refractive index aerogel. The pinhole-dried aerogels are generally more transparent than the original non-shrink ones. Moreover, the most transparent aerogel can be produced by the pinhole drying method in the middle range ($n$ $\thicksim$ 1.06) as shown by the upper solid line. There are three pinhole-dried aerogel groups which are shown by the lines in Fig. \ref{fig:fig5}. Each group has roots in certain original non-shrink aerogels with different refractive indices. The solid and dashed lines indicate pinhole-dried aerogels synthesized by using DMF and methanol solvents, respectively. Although transmission lengths of the pinhole-dried aerogels synthesized by using conventional methanol solvent shown by the dashed line are not always the best ones, the aerogels have the advantage that they can be successfully produced in a short pinhole drying period compared to DMF solvent. 

%
%
%
%

\section{Beam test}

To evaluate pinhole-dried aerogels as a Cherenkov radiator, newly produced aerogels were irradiated with 2 GeV/c electron beams at the KEK Fuji test beam line in 2008. A prototype aerogel RICH counter with a $2\times 3$ array of 144 channel multi-anode hybrid avalanche photon detectors (HAPD) was built and electron beams were tracked by two MWPCs arranged up and down stream of the RICH counter. The detail of the experimental set up is found in Ref. \cite{example7}.

\begin{figure}[t]
\begin{center}
\includegraphics*[scale=0.30]{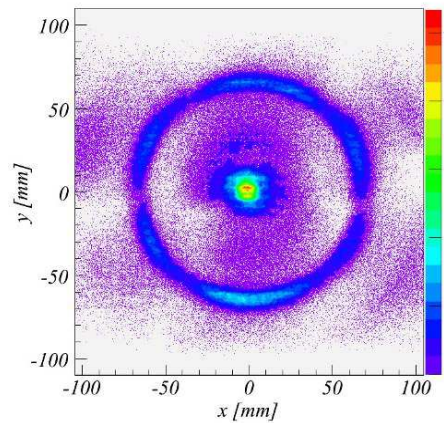}
\end{center}
\caption{\label{fig:fig6}
Accumulated hit distribution of Cherenkov photons at the photon-detection plane. For each hit, the difference between the beam position and the photon detector hit position is plotted. The arrangement of the photon detectors ($2\times 3$ array of HAPDs) at 10 mm intervals led to the Cherenkov ring image with some gaps.
}
\end{figure}

Fig. \ref{fig:fig6} shows an observed clear Cherenkov ring image of the pinhole-dried aerogel sample with $n$ = 1.198. The measured Cherenkov angles of pinhole-dried aerogels were consistent with the expected values from the refractive indices measured by the laser Fraunhofer method. In the case that the thickness of aerogel radiators is 1 cm, the $N_{pe}$ of the aerogel with $n$ = 1.198 was 9.8 while that of an reference aerogel with $n$ = 1.066 was 3.9.

%
%
%
%

\section{Conclusion}

We have succeeded in developing hydrophobic silica aerogel with refractive indices over a range of 1.0026-1.26. Especially, the pinhole drying method enables to obtain highly transparent aerogels with high refractive indices up to $n$ = 1.26. It was confirmed that pinhole-dried aerogels can be used as a Cherenkov radiator by the beam test. On the other hand, extremely low density aerogel ($\rho $ = 0.01 g/cm$^3$) will be utilized as a cosmic dust capturer in the sample return experiment on the JEM/ISS.

%

\section*{Acknowledgments}

We would like to thank the Belle and TANPOPO colleagues for their assistance. This work is partially supported by JSPS Grant-in-Aid for JSPS Fellows (Grant no. 07J02691).

%
%
%
%
%
%

\end{document}